\newcommand{\beq}{\begin{equation}}
\newcommand{\eeq}{\end{equation}}
\def\etal{{\it et al. \/}}
\def\eg{{\it e.g.\/}}
\def\ie{{\it i.e.\/}}
\def\rosat{{\sl ROSAT \/}}
\def\asca{{\sl ASCA \/}}
\def\pros{{\sl PROS \/}}
\def\simless{\mathbin{\lower 3pt\hbox
     {$\rlap{\raise 5pt\hbox{$\char'074$}}\mathchar"7218$}}} 
\def\simgreat{\mathbin{\lower 3pt\hbox
     {$\rlap{\raise 5pt\hbox{$\char'076$}}\mathchar"7218$}}} 

\documentstyle[emulateapj]{article}

\submitted{revised version submitted to ApJ 04 Dec 2000}

\lefthead{SULKANEN AND BREGMAN}
\righthead{Are there ADAFs in Nearby Giant Elliptical Galaxies?}

\begin{document}

\title{Are there ADAFs in Nearby Giant Elliptical Galaxies?}

\author{Martin E. Sulkanen\altaffilmark{1,2}}
\affil{X-Ray Astronomy Group, Space Sciences Laboratory, \\
 NASA/Marshall Space Flight Center, Huntsville, AL 35812 \\
 msulkanen@astro.lsa.umich.edu \\}

\altaffiltext{1}{Rackham Visiting Scholar, University of Michigan.} 
\altaffiltext{2}{present address: Department of Astronomy,
    University of Michigan, Ann Arbor, MI 48109-1090}

\author{Joel N. Bregman}
\affil{Department of Astronomy, University of Michigan,
Ann Arbor, Michigan 48109-1090 \\
jbregman@astro.lsa.umich.edu \\}

\begin{abstract}

Recently it has been suggested that the hard x-ray power-law tails exhibited 
in the \asca spectra of several nearby giant elliptical galaxies 
are caused by low-radiative efficiency accretion flows (ADAFs) 
onto a massive black hole in the centers of the galaxies. The estimated 
fluxes from these central ADAFs as derived from the \asca x-ray 
spectral analyses indicate that they may be visible as x-ray point 
sources. We analyze archival \rosat HRI images of three such galaxies, 
NGCs 1399, 4636, and 4696, to determine whether point x-ray 
sources consistent with \asca-derived fluxes are present in the centers
of the galaxies. We find that the upper limit for the flux of a central
point x-ray source in each of these galaxies as determined by the 
\rosat HRI data is only marginally consistent with the predicted flux
inferred from the ASCA spectra. We suggest that although a central 
point source, such as a ADAF associated with a massive black hole, 
cannot be completely ruled out, they can only account for a fraction
of the flux of the observed hard x-ray power law components.

\end{abstract}


\section{INTRODUCTION}

The presence of both massive ($10^9 M_\odot$) black holes
in the centers nearby elliptical galaxies 
(\markcite{KoRi95}Kormendy and Richstone 1995) and  
extensive atmospheres of $kT \sim$ keV gas (\markcite{TrFa86}Trinchieri,
Fabbiano, and Canizares 1986)
suggest that these galaxies should also possess central 
point x-ray sources of luminosities of $L_x \sim 10^{45}$ erg sec$^{-1}$
associated with the accretion of gas onto the black hole. 
In reality these galaxies exhibit x-ray luminosities $< 10^{-3}$
of this magnitude, including the contribution from the hot galactic 
atmosphere (\markcite{Pell99}Pellegrini 1999). This fact has 
motivated the development of scenarios
where accretion onto the central black hole occurs via  
``advection-dominated accretion accretion flows'' (ADAFs)
(\eg \markcite{NaYi95} Narayan and Yi 1995; 
\markcite{Abrm95} Abramowicz \etal 1995; 
\markcite{DiMa00} DiMatteo \etal 2000).
By means of poor collisional coupling between very hot ions and radiatively
efficient electrons, and the presumption of a coexisting wind or 
outflow (\markcite{BlBg99}Blandford and Begelman 1999), 
such scenarios result in accretion onto the black hole at a small fraction
of the Bondi (\markcite{Bndi52}1952) rate, producing a correspondingly small
x-ray luminosity. Furthermore, hard power-law 
components (photon index $\Gamma \sim 0.6 - 1.5$) 
detected in the \asca x-ray spectra of six nearby giant elliptical 
galaxies with radio-quiet cores have been suggested to 
be produced by an ADAF atmosphere around their central black holes
(\markcite{AlDF00}Allen, DiMatteo, and Fabian 2000, hereafter ADF2000). 
These spectra are for the entire galaxies, and although contamination from 
from galactic x-ray binaries is possible, ADF2000 suggest that the 
power-laws exhibited in the spectra are significantly harder 
than that produced by x-ray binaries or from standard AGN. If 
these hard power-laws really are black hole-ADAF signatures, then based 
on the ADF2000 estimates for their photon indices and normalizations 
they could appear as relatively strong point sources in the \rosat HRI: 
although the ADAF mechanism can supress a very large x-ray brightness, in
principle it can still be detected. Some of these galaxies offer
circumstantial evidence of nuclear activity. The HRI analysis of M87 
by Reynolds \etal (\markcite{Reyn96}1996) and Harris \etal 
(\markcite{Harr98}1998) combined with the ADF2000 spectral analysis 
make a plausible case
for the association of M87's hard power-law tail with a massive black
hole  ADAF (see ADF2000). 
M87 also posesses a prominent x-ray and radio jet, thus energetic
activity associated with a central source is clearly indicated.
The cores of NGC 1399, 4636, and 4696 are much weaker in radio power than 
M87 (\markcite{Slee94}Slee \etal 1994).

We consider whether the preponderance of the 
observed hard x-ray emission in some of the galaxies discussed in 
ADF2000 should be attributed to a massive 
black hole-ADAF. DiMatteo \etal (\markcite{DiMa99}1999; hereafter D99) 
determined upper limits for three of these galaxies, NGCs 4472, 4636,
and 4649; these were comparable to the \asca fluxes in the case of 
NGCs 4472 and 4649, but in the case of NGC 4636 the HRI upper limit 
was a factor of two lower than the \asca measurement. In
this article we present analysis of \rosat HRI images of 
additional ADAF candidates from the list of the 
ADF2000 galaxies, NGCs 1399, 4696, as well as NGC 4636.  
We find that the upper limit for the flux of a central
point x-ray source in each of these galaxies as determined by the
\rosat HRI data is either smaller than the predicted flux
inferred from the ASCA spectra, or only marginally consistent. 
Our result for NGC 4636 is comparable
to that determined by D99,\ie, lower than the \asca inferred
flux by a factor of $\simeq 1.4$.  Thus, if ADAF sources 
associated with central black holes exist in these galaxies, they can
only account for only a fraction of the hard power-law x-ray emission observed
in the galaxies' \asca spectra.  

\section{OBSERVATIONS}

The observations were obtained from the HEASARC \rosat data archive, 
and are summarized
in Table 1. The observations were subjected to the ``dewobbling''
recipe described in the \pros User Guide (\markcite{PRG98}1998) 
in order to improve the spatial resolution of the HRI image. This
procedure groups the HRI photons by phase bins of the telescope's 
$402$ sec wobble period, determines a separate image for each of
these phase bins, and then reassembles the images to a common
image centroid. We have checked whether this scheme recovers the
\rosat/HRI point source response function: in the case of NGCs 1399
and 4636, point sources within a few arcmin of the HRI field center 
appear to have a corrected FWHM of $\simeq 6$\arcsec. Therefore we
have used this FWHM for the convolution of the model galactic
surface brightness, including a central point source (see \S 3).
In the case of NGC 4696, there is no point source of sufficient strength
within the HRI image field to check the recovery of the HRI resolution,
so we have assumed a 6\arcsec FWHM for model fitting. 
The effective observation times are calculated
from the good time intervals covered by the list of wobble phases used to
compose the dewobbled image.     

\begin{table*}
\begin{center}
\begin{tabular}{ccc}
Target &\rosat Seq. ID &Exposure \\
NGC&  &(ksec)\\ 
\tableline
1399 &rh600831a01 & 76699\\
4636 &rh600218a01$+$a02 & 11813\\
4696 &rh700320a01 & 13822\\
\end{tabular}
\end{center}
\caption{The log of ROSAT analyzed observations for each galaxy}
\end{table*}
  
\section{SPATIAL ANALYSIS AND RESULTS}

Using the dewobbled images, we extract azimuthally-averaged 
counts per HRI pixel in 24 $2.5$\arcsec wide
annulae centered on the galaxies. The x-ray centers are determined by
the peak of x-ray contours of the surface brightness of the galaxy core.
We note that in the cases of NGCs 1399 and 4636 the \rosat x-ray position
agrees within $\simeq 2$\arcsec of their optical centers (the putative
central black hole position), while the x-ray peak for NGC 4696
is within $\simeq 6$\arcsec of both the HST position for the nucleus and
the location of the nuclear radio source (Sparks, private communication;
\markcite{ODea94}O'Dea \etal 1994). This may be consistent with the
\rosat pointing accuracy, however if the x-ray peak is not associated with
the true galactic nucleus, the upper limit to the flux from an ADAF associated
with a nuclear black hole is much smaller than the value we determine
from our model that we describe below. From each annulus we subtract a 
background determined from a 5\arcsec wide annulus at radius of $\sim$
12\arcmin. The pixel brightness in a given annulus is assigned to the 
mean radius of the annulus. 
Background point sources are removed where necessary.
The datapoints are fit to a $\beta$-profile model
for the extended galactic x-ray surface brightness, 
$S_X = S_{X0} (1 + r^2/r_c^2)^{3\beta - 1/2}$, and a central 
central point source, both convolved with
the dewobbled \rosat HRI point source response function 
of 6\arcsec (\markcite{BrBr99}Brown 
and Bregman 1999). This composite model is a simple representation
of the essential question that we wish to answer: to what extent must
a central point source be included with extended emission to describe
the galactic surface brightness? Note that the $\beta$-profile model
may be less than ideal to describe the extended emission - it cannot
account for asymmetry and substructure (\eg NGC 4696) - however,
we are not attempting to find the best model to describe the global
x-ray emission, but a measure as to the point-source like behavior
of the center compared to a flat emission ``core''. Given the core-like
behavior of the emission at small radii, and the 
power-law like behavior at large
radii, the $\beta$-profile model is a reasonable proposition as a gross
description of the extended emission of the galaxies. 

The $\beta$-profile + point source model best fit results, 
and the 99\% confidence upper limits to the count rates from 
a central point source in each galaxy are listed in Table 2. 
These are determined by increasing the magnitude of fixed point 
source until the fit for the remaining $\beta$-profile parameters yields a 
$\Delta \chi^2 \geq 6.64$ above the best fit.
The maximum point source strength determined in precisely the same manner
for the undewobbled images is substantially
lower for all three galaxies, by factors of $1.5 - 7$.
Plots of the surface brightness for all three galaxies, along with
the profiles of the best fit models and those with the upper limit
of central point source brightness, are given in figures 1-3.  

\begin{table*}
\begin{center}
\begin{tabular}{cccc}
Target &$\chi^2$/dof (best fit)&HRI Central Point Source Rate &$N_H$ (PSPC/Galactic)\\
NGC  & &sec$^{-1}$ &$10^{20}$ cm$^{-2}$\\
\tableline
1399 &$30.4/20$&$\leq 4.9 \times 10^{-3}$&$1.39$\\
4636 &$41.7/20$&$\leq 4.8 \times 10^{-3}$&$1.90$\\
4696 &$25.5/20$&$\leq 2.0 \times 10^{-3}$&$10.15$\\
\end{tabular}
\end{center}
\caption{The beta-profile + central point source best fit results, and the
99\% confidence level upper limits for \rosat HRI count rates.
Also included are the galaxies' absorption column densities, 
taken to be the larger
of the measured values of Stark \etal (1992) and the
best-fit values for \rosat PSPC spectra analyzed by Davis and White 
(1996).} 
\end{table*}

Figures 4-6 shows the 2-10 keV flux and photon index $\Gamma$ for the
\asca power-law components determined
by ADF2000 (including 90\% confidence limits on a single interesting parameter) 
compared to our upper limits for the flux determined from the
HRI count rates for a central point source, as a function
of $\Gamma$. These PIMMS estimates are shown 
for the column absorption densities given by the maximum value of either
Stark \etal (\markcite{Star92}1992) or the best-fit values derived
from analysis of \rosat PSPC spectra (Davis and White {\markcite{DaWh96}1996;
see Table 2).

\section{Discussion and Summary}

We note that for all three galaxies the optimal values for the \asca 
flux and photon index fall above our HRI upper flux limits, including
effects of absorption, although the large uncertainties in photon index
allow for marginal consistency between the \asca values and the upper
flux limits (\eg NGC 1399). 
ADAF models for the hard x-ray emission component suggest
a photon index of $\Gamma \sim 1.4$, and the ensemble average of \asca results
for the putative ADAF power law components is 
$\Gamma \simeq 1.2$ (ADF2000). Our results indicate that for NGCs 1399 and 4696
photon indices $\Gamma > 1$ are ruled out unless the flux from a putative
ADAF is a fraction ($\sim 0.40 - 0.70$) of the \asca power-law component.
In the case of NGC 4636, our result approximately agrees with an 
earlier analysis by D99, namely that at the optimal photon index 
the HRI upper limit for the flux is significantly lower (here a factor
of $\simeq 1.4$) than the measured \asca
flux, including the effects of measured absorption. Unlike the other 
two galaxies, photon indices $\Gamma < 1$ for the \asca power-law component
are ruled out. A ``universal'' photon index of $\Gamma \simeq 1.2$ to
describe the bulk of the hard x-ray emission from an ADAF source in 
all three of these galaxies is not consistent with the HRI upper flux
limits. In the case of NGC 4636, the larger photon index may be consistent
with the presence of a significant contribution in the power-law component 
from other sources, such as binary x-ray sources. ADF2000 noted that 
extrapolation of the $L_X/L_B$ relation of Fabbiano and Trinchieri 
(\markcite{FaTr85}1985) measured for irregular and spiral galaxies
of the power-law components of  NGCs 1399 and 4636 was consistent with
substantial contribution from x-ray binaries, stars, and an AGN. 
In the case of NGC 1399 there is also some indication that the hard 
component may be spatially distributed. Jones \etal (\markcite{Joet97}1997)
found that the PSPC spectrum for the region outside a $3.5$\arcmin core 
of the galaxy was best fit with a $\sim 1$ keV and a harder $2-3$ keV 
two-temperature plasma model, while the core was well described by
a single temperature plasma of $\sim 1$ keV. This outer region overlaps
with the 6\arcmin circular extraction region for the \asca GIS used in
the analysis of this galaxy by ADF2000.

ADF2000 also suggest that the flat photon indices measured for
these sources by \asca are central ADAFs with intrinsic photon
indices of $\Gamma \sim 1.4$, flattened by anomalously high absorption
along the cores of these galaxies of $N_{\rm H} \simgreat 
{\rm few} \times 10^{21}$ cm$^{-2}$. However, such an excess is not indicated
from \rosat PSPC observations of Davis and White (\markcite{DaWh96}1996),
nor is excess absorption is observed by the \rosat PSPC in giant 
elliptical galaxies containing cooling inflows located at low 
galactic column, such as NGC 1399 (\markcite{Joet97}Jones \etal 1997;
\markcite{Davi94}David \etal 1994). Buote (\markcite{Buot00}2000) 
has suggested a resolution of the disagreement between the \asca 
and \rosat column estimates by oxygen absorption at $0.4 - 0.8$ keV 
caused by a warm component of gas in the core of NGC 1399, associated 
with mass dropout in the cooling inflow. As all three of the galaxies 
we consider here have cooling inflows, we leave it as a possibility 
that this effect has supressed the brightness of their central ADAFs 
as observed by the \rosat/HRI. 
  
We present the HRI spatial analysis of the central regions of three
nearby giant elliptical galaxies in which hard power-law components
have been detected in the \asca x-ray spectrum. We find the overlap 
of error regions for the \asca power-law flux and photon indices 
with our HRI-determined 99\% confidence upper flux limits for a 
central point source is either marginal or small for all three galaxies. 
Thus associating the entirety of this hard x-ray component with 
a central x-ray source (like an massive black-hole ADAF) is unlikely,
unless anomalously high absorption, possibly related to the cooling
inflows in the cores of each of these galaxies, absorbs a substantial
amount of the ADAF x-ray flux in the \rosat passband. 
Otherwise, if we require that the range of possible values
of $\Gamma$ as determined by ADF2000 is consistent with our upper limits
for the galaxies' flux, then the flux
attributed to the central source can only be a fraction
of the measured \asca flux for the observed power-law component. 

We thank Beth Brown for providing her $\beta$-profile fitting routine.

This research has made use of data obtained through the High Energy
Astrophysics Science Archive Research Center Online Service, provided
by NASA/Goddard Space Flight Center.

M.E.S. also thanks NASA's Interagency Placement Program, 
the University of Michigan Department of Astronomy, 
and the University of Michigan Rackham Visiting Scholars Program.

}

\begin{figure}
\plotone{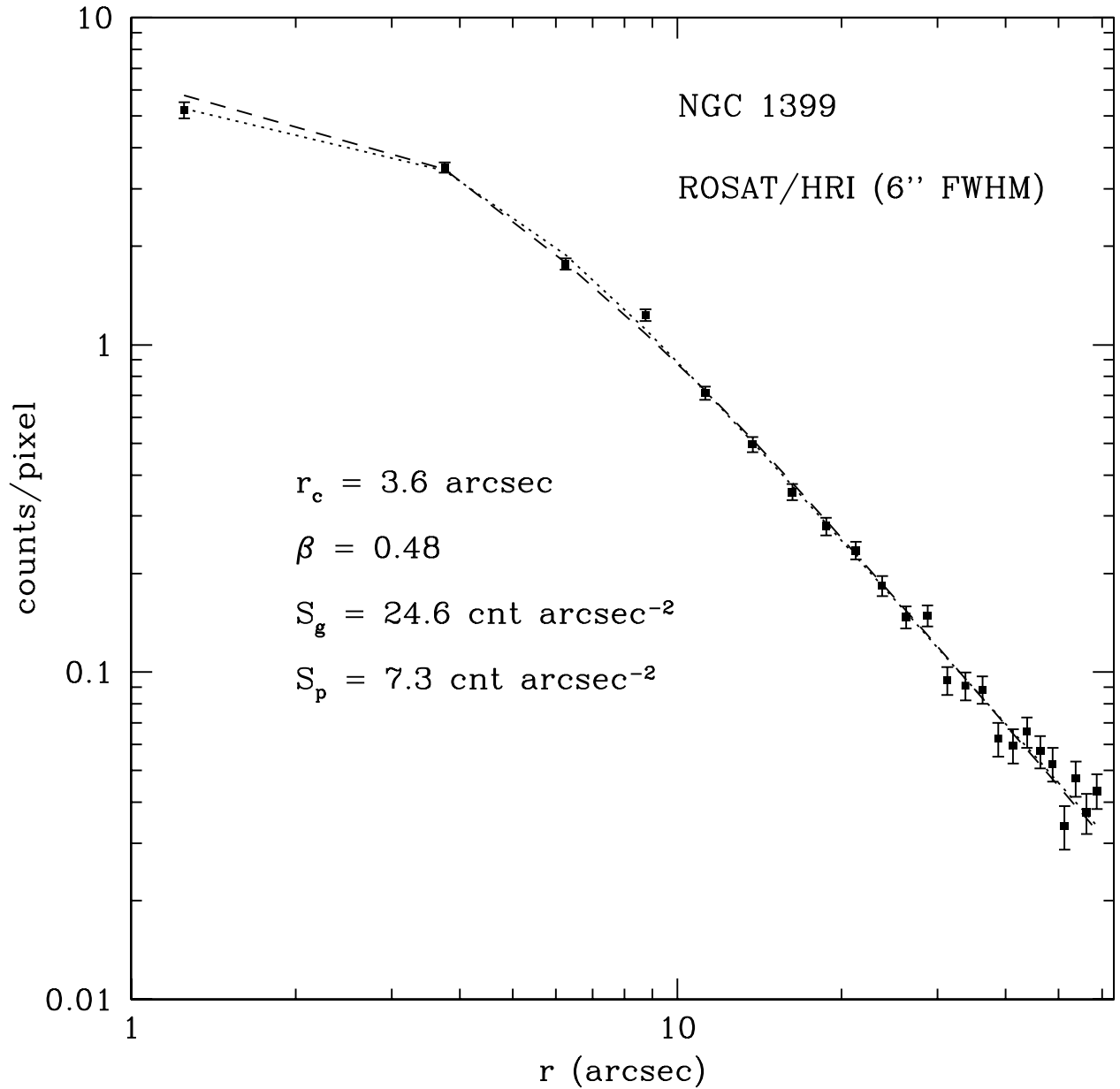}
\caption[]{Plot of the radial surface brightness for 
NGC 1399, along with the profiles of the best fit of a $\beta$-profile +
central point source model (dotted line; best-fit values of parameters
given in legend), and that with the upper limit
of central point source brightness (dashed line).}
\end{figure}

\begin{figure}
\plotone{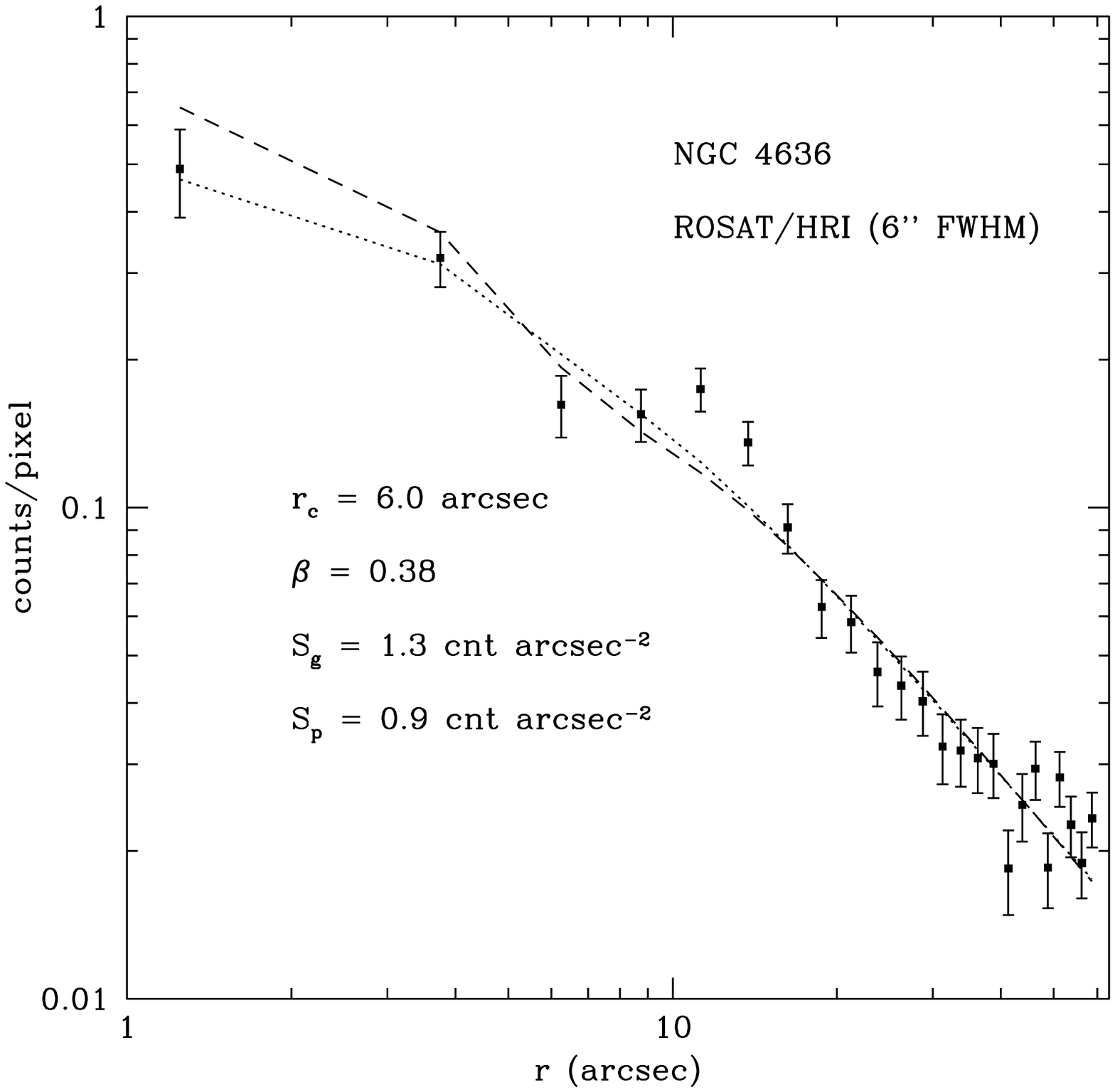}
\caption[]{Plot of the radial surface brightness for 
NGC 4636; lines and legend identical to figure 1.}
\end{figure}

\begin{figure}
\plotone{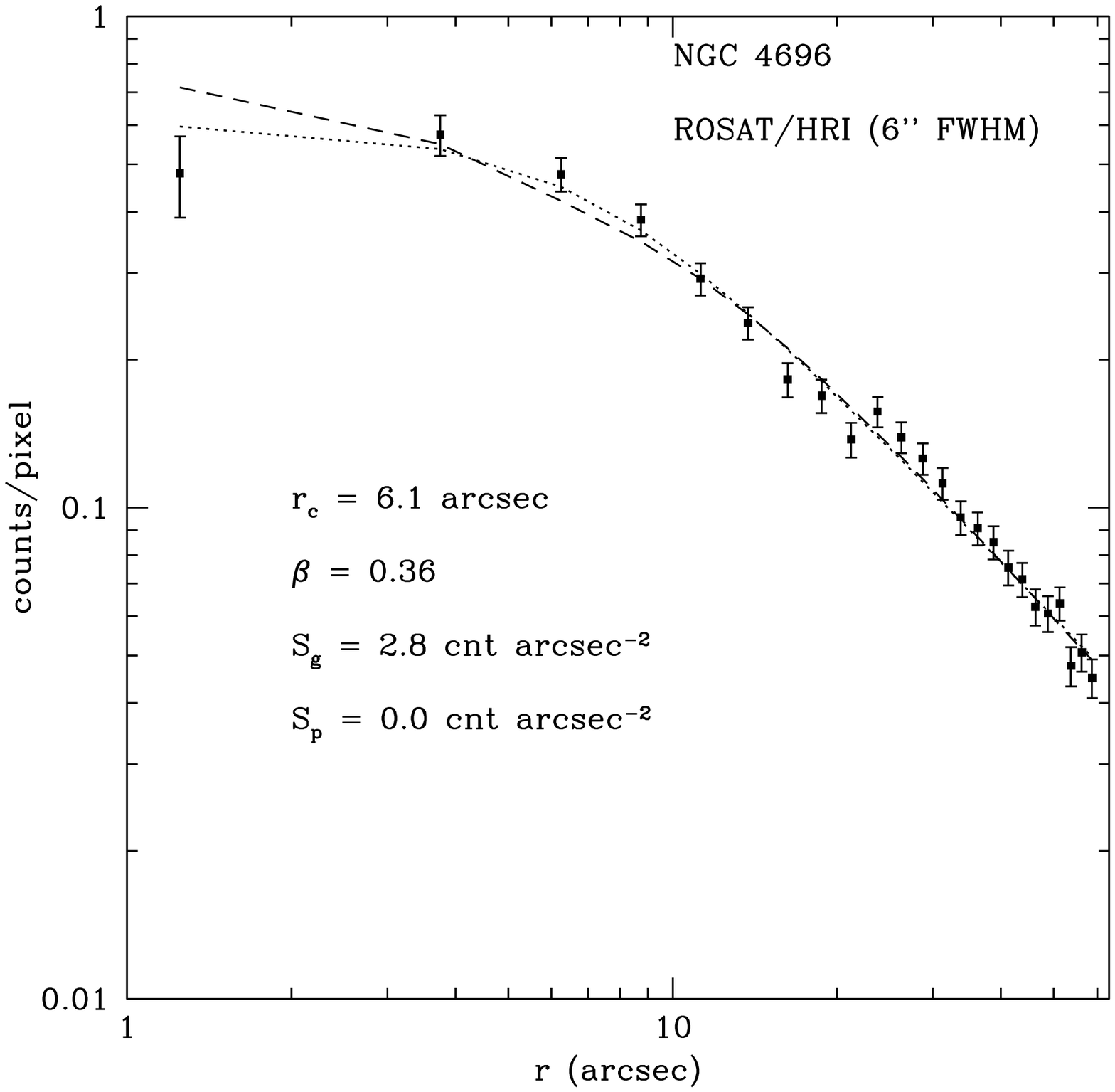}
\caption[]{Plot of the radial surface brightness for 
NGC 4696; lines and legend identical to figure 1.}
\end{figure}

\begin{figure}
\plotone{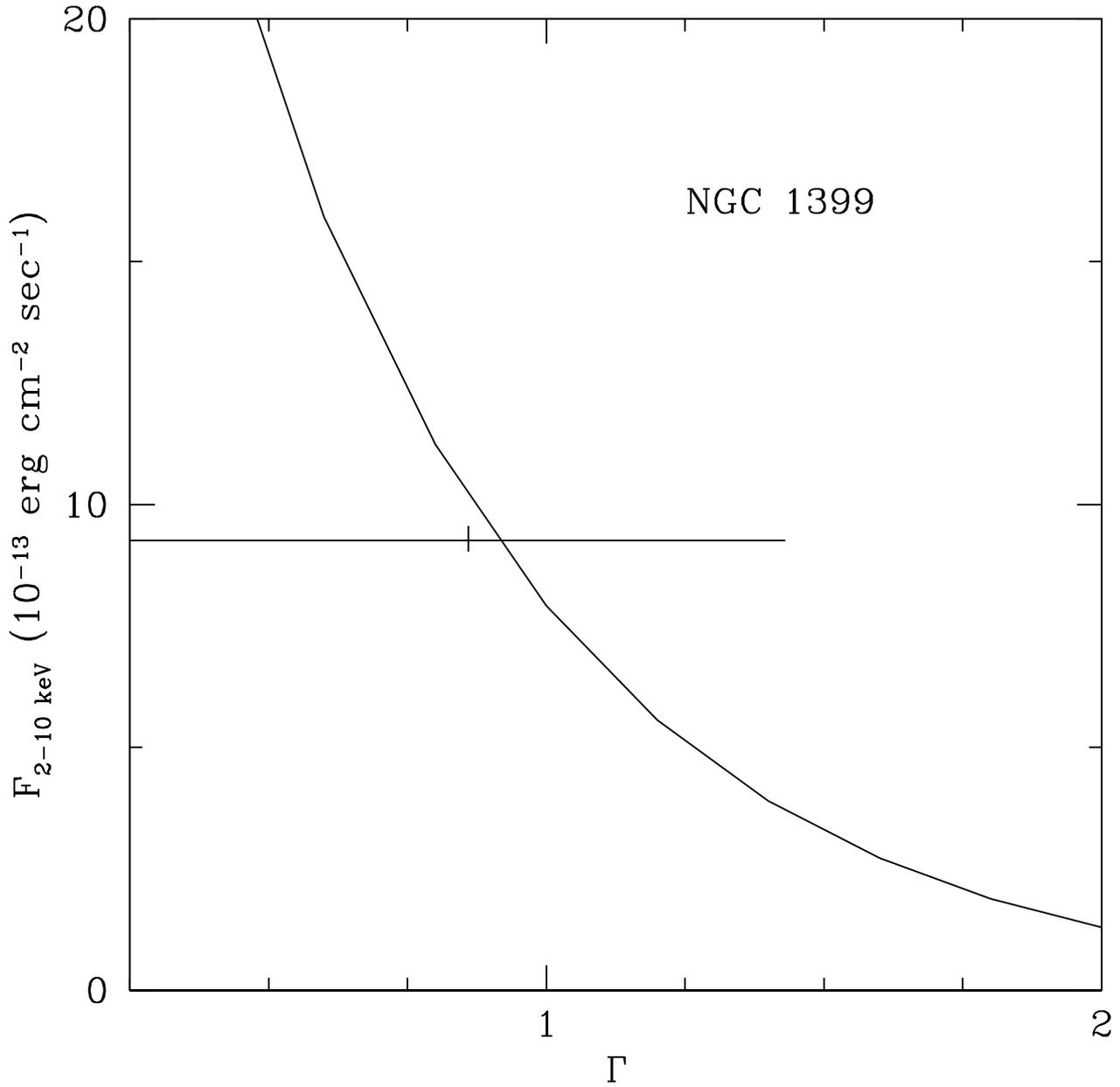}
\caption[]{The curve indicates the upper limit to the 2-10 keV 
flux for a central point source
in NGC 1399, assuming a power-law spectrum, derived from the 
HRI count rate upper limit (99\% confidence limit), and includes
the value of absorption column 
taken from Table 2. The abcissa is the
power-law photon index $\Gamma$. 
The data point plotted is the best fit value for flux and $\Gamma$ determined
from the \asca spectrum analyzed by ADF2000, with 90\% confidence error bars.}
\end{figure}

\begin{figure}
\plotone{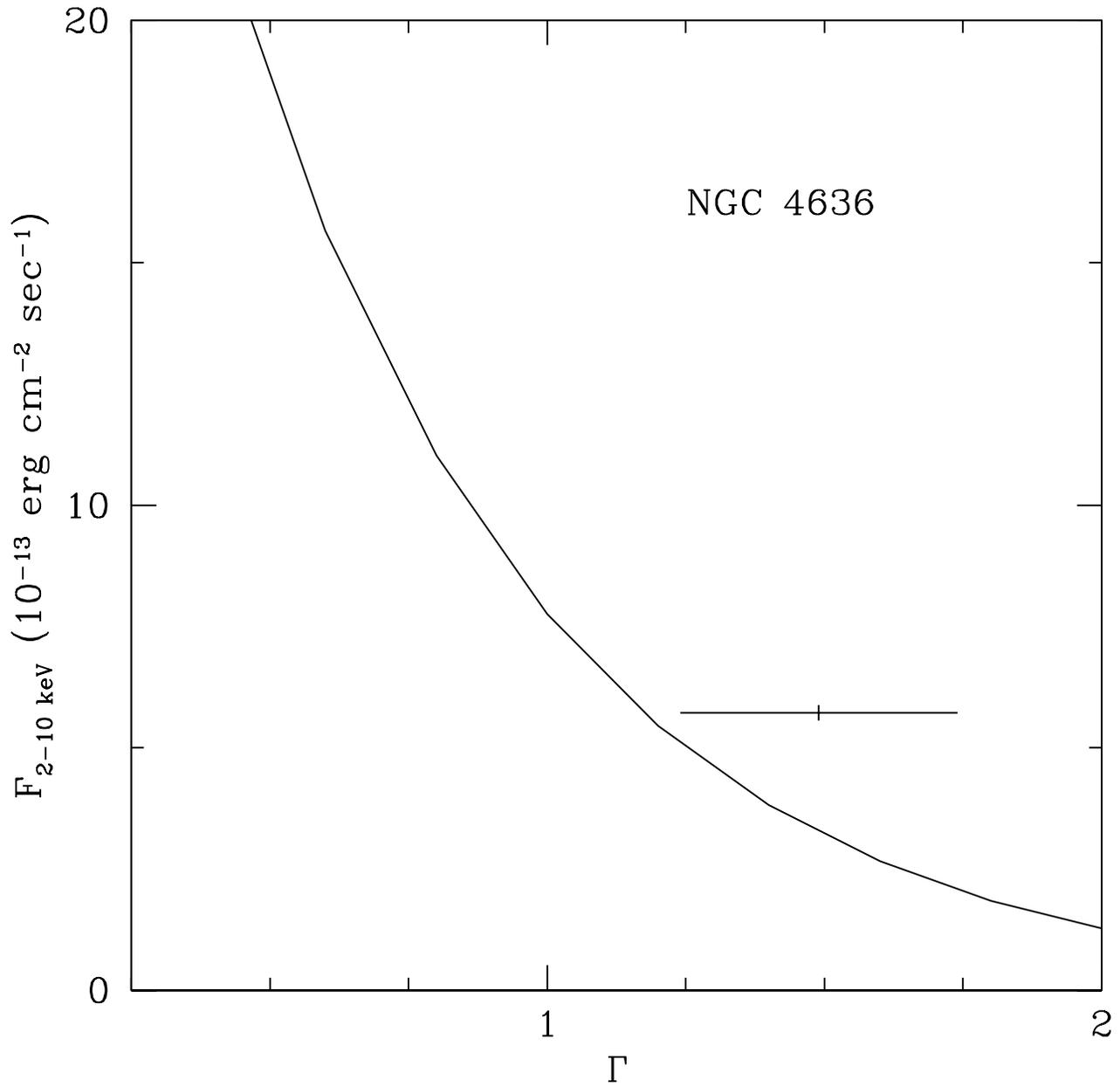}
\caption[]{Plot identical to Fig. 4 for NGC 4636; with absorption
taken from Table 2.}
\end{figure}

\begin{figure}
\plotone{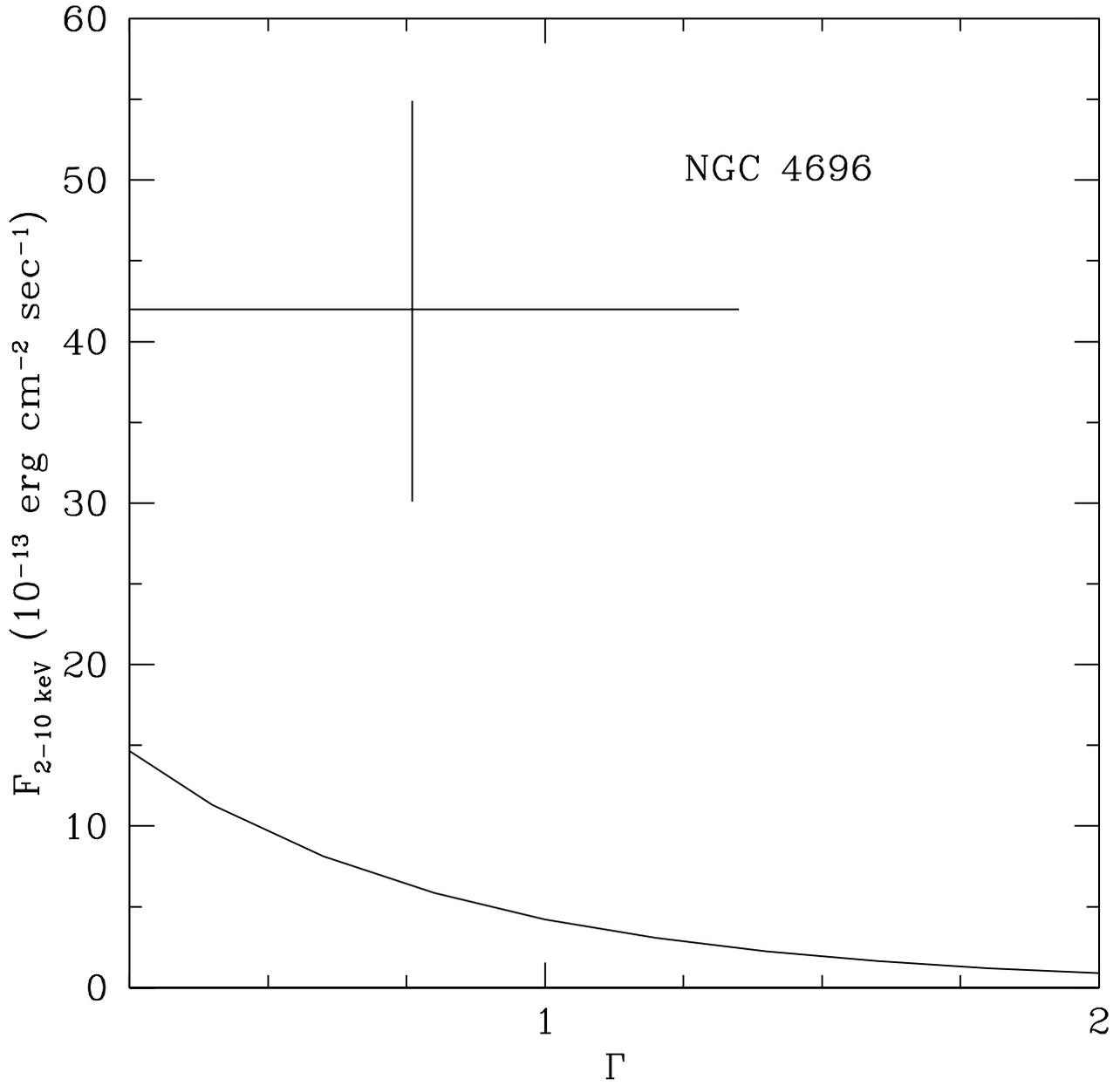}
\caption[]{Plot identical to Fig. 4 for NGC 4696.}
\end{figure}

\end{document}